# Water Vapor: An Extraordinary Terahertz Wave Source under Optical Excitation

Keith Johnson[a,b], Matthew Price-Gallagher[b], Orval Mamer[c], Alain Lesimple[c], Clark Fletcher[b], Yunqing Chen[d], Xiaofei Lu[d], Masashi Yamaguchi[d], and X.-C. Zhang[d]

[a] *Massachusetts Institute of Technology, P.O. Box 380792, Cambridge, MA 02238-0792,* [b] *HydroElectron Ventures Inc.,1303 Greene Avenue Suite 102, Westmount, QC, Canada, H3Z 2A7,* [c] *Mass. Spec. Unit, 740 Dr. Penfield, Suite 5300, McGill Univ., Montreal, QC Canada, H3A 1A4,* [d] *W.M. Keck Laboratory for Terahertz Science, Center for Terahertz Research, Rensselaer Polytechnic Institute, Troy, NY 12180, USA*

**Abstract**

**In modern terahertz (THz) sensing and imaging spectroscopy, water is considered a nemesis to be avoided due to strong absorption in the THz frequency range. Here we report the first experimental demonstration and theoretical implications of using femtosecond laser pulses to generate intense broadband THz emission from water vapor. When we focused an intense laser pulse in water vapor contained in a gas cell or injected from a gas jet nozzle, an extraordinarily strong THz field from optically excited water vapor is observed. Water vapor has more than 50% greater THz generation efficiency than dry nitrogen. It had previously been assumed that the nonlinear generation of THz waves in this manner primarily involves a free-electron plasma, but we show that the molecular structure plays an essential role in the process. In particular, we found that THz wave generation from $H_2O$ vapor is significantly stronger than that from $D_2O$ vapor. Vibronic activities of water cluster ions, occurring naturally in water vapor, may possibly contribute to the observed isotope effect along with rovibrational contributions from the predominant monomers.**

Water has traditionally been anathema to spectroscopy in the terahertz (THz) region of the electromagnetic spectrum. Water vapor shows a dense collection of extremely strong absorption resonances within the far infrared [1] and must be purged from the spectrometer for most measurements. Despite the ubiquity of this antagonist, intense interest has been focused on the THz region due to its potential in wide ranging disciplines from medicine to security to pure science. With recent advances in the optical production of intense THz fields, THz science is on the verge of crossing over from the heavily studied realm of linear far infrared spectroscopy into nonlinear spectroscopy [2]. A promising candidate to lead this transition is the generation of pulsed THz waves in air via four-wave mixing [3-7]. Previous work has attributed the physical mechanism to a free electron process in the laser induced plasma formed at the focus of an amplified Ti:Sapphire laser [4-8]. However, in this letter we show by a systematic comparison between different gases that the signal is strongly dependent on the molecular or atomic species and that of the samples tested the old nemesis of THz laboratories, water vapor, generates the strongest THz radiation on a per-molecule basis. This provides a THz wave source with unlimited availability and offers a unique window into the nonlinear responses of highly excited molecular states. Significantly, we also have found that THz wave generation from $H_2O$ vapor is substantially stronger than that from $D_2O$ vapor, suggesting that a quantum mechanical treatment of molecular vibrational dynamics may more closely reflect the dominant physical mechanism. Along with rovibrational contributions from the predominant monomers, this may include the vibronic activities of naturally occurring water clusters [9-12]. Since water vapor is Earth's most potent greenhouse gas, its THz emission as well as its infrared absorption spectrum may help elucidate the possible relevance of atmospheric water clusters to global warming [13].

We have investigated THz wave generation from water vapor at pressures from mTorr to over 800 Torr with two complementary methods: water vapor in a cell (low pressure, ambient temperatures, and high pressure) *and* from a nozzle jet in a vacuum chamber (middle pressure and semi-high pressure). Fig. 1 schematically illustrates the experimental arrangement. A tabletop Ti: Sapphire regenerative amplifier provides laser pulses with 100 fs duration, 1 mJ pulse energy, 1 kHz repetition rate, and 800 nm central wavelength. Intense fundamental laser pulses ($\omega$) and their second-harmonic pulses ($2\omega$) are focused into a gas (contained in a cell or emitted from a jet nozzle) with a 100 mm focal length lens. The second-harmonic pulses are generated by a 100-μm thick type-I beta barium borate (BBO) crystal. Fig. 1 inserts show the detail of the plasma generating regions using the cell or pulsed jet nozzle. After the mixing process occurs at the focus, the generated THz wave is collimated and refocused by a pair of off-axis parabolic mirrors, and detected via electro-optic sampling in either a 2.5 mm thick <110> oriented ZnTe crystal (higher dynamic range) or a 0.1 mm thick <110> GaP crystal (broader bandwidth). The entire system is placed within a chamber which can be evacuated or purged with dry nitrogen.

About 20-40 ml liquid water was put into a flask, and frozen by liquid nitrogen. A vacuum pump was used to evacuate the flask and cell to 1-5 mTorr, and then the valve connected to the flask was closed. The flask was warmed and the ice melted to provide pure water vapor. With a gas dilution system and a vacuum pump, a specified pure vapor (0.1-23 Torr) can be input into the gas cell and the pressure inside can be precisely controlled. As a comparison, pure nitrogen was tested as THz emitter under the same experimental conditions.

The glass gas cell has a diameter of 30 mm and a length of 50 mm. The front window is a 150 μm thick fused quartz plate; the rear window is made of high-density polyethylene. With the cell connected to a gas diluting system, gas is introduced into or removed from the gas cell, and the pressure inside is precisely controlled with an accuracy of 1 mTorr in the 1 mTorr-10 Torr range and 0.1 Torr in the 10-1000 Torr range.

Two major issues associated with the vapor cell should be considered: one is the optical dispersion of water vapor between ω (800 nm) and 2ω (second harmonic beam, 400 nm) from entrance window to the plasma spot (2-3 cm). Another one is THz wave absorption by water vapor between the plasma and exit window inside the cell (2-3 cm). Under low pressures (23 Torr or below), the absorption of water vapor accounts for less than 1% of emitted THz strength in the plasma according to Beer's law, which can be ignored. Moreover, the dispersion of water vapor below 23 Torr is small since Δn (the phase difference between 800 nm and 400 nm) is small at lower vapor pressure. However, at higher water vapor pressure (>100 Torr), these dispersion effects must be considered.

Using a gas nozzle jet, both optical dispersion and THz wave absorption in the vapor cell can be avoided. The measurements were performed using a pulsed gas jet in order to minimize the effects of dispersion and re-absorption in the gas being tested. The gas jet is synchronized with the laser, delivering 400 μs pulses at a repetition rate of 100 Hz. Since the optical focus is very close to the jet opening (within 1 mm), the pressure reaches a steady state within 200 μs, indicating that the maximum gas density at the focus is approximately equal to the backing density. The timing between the gas and laser pulses is adjusted electronically such that their maxima overlap. The temperature was measured by a thermocouple located in the gas tubing directly behind the valve opening. This allowed more precise pressure control since the temperature could be monitored to ensure that the water vapor was saturated, which is necessary to avoid condensation. The disadvantage in using a nozzle is that the exact pressure and temperature value at the plasma location have to be approximated.

Fig. 2 compares the THz generation efficiency, defined as THz amplitude / (pump energy x molecules), of water vapor, with that of several noble gases and nitrogen gas. Terahertz pulse generation in the laser-induced plasma from a series of noble gases, He, Ne, Ar, Kr, and Xe was systematically investigated [14]. Experimental results reveal that terahertz generation efficiency of these noble gases increases with decreasing ionization potential. Water vapor, viewed simplistically as noninteracting $H_2O$ monomers with the ionization potential indicated in Fig. 2, does not follow the general trend with the noble gases at 20 Torr. However, the ionization potentials of *water clusters*, a natural component of water vapor, are significantly lower than that of the monomer, as indicated for the protonated cluster, $(H_2O)_{21}H^+$, and follow more closely the general trend of the noble gases. The point here is that water vapor has more than 50% greater THz generation efficiency than dry nitrogen, with its ionization potential between Xe and Kr.

Fig. 3a shows the pressure dependence of THz wave generation for $H_2O$, $D_2O$, and $N_2$. This measurement is performed in both the cell and jet. For pressures less than 20 Torr, the data were taken from the cell, while at higher pressure (>20 Torr), the data were taken from the pulsed nozzle. $H_2O$ generates about 5% to 10% more THz wave radiation than $D_2O$ between 100 mTorr and 1000 Torr. Fig. 3b plots both the frequency and pressure dependence of THz wave generation in water vapor. To clarify whether THz generation from water vapor is enhanced by four-wave mixing, ponderomotive force, etc., emission from single wavelength excitation at 800 nm was compared with that of excitation at two wavelengths (800 nm + 400 nm). The THz amplitude generated in 22.5 Torr water vapor at 800 nm is two orders weaker that of 800 nm + 400 nm with the same total optical power.

Our observations are: (1) Water vapor produces stronger THz emission than nitrogen at the same pressure and power; (2) The threshold of pump energy for THz generation from water vapor is about 100 µJ at 200 mTorr; (3) The strong THz generation from water vapor is related to a four-wave mixing process; (4) No specific molecular signatures were found in the THz emission spectrum of water vapor; (5) THz wave generation from $H_2O$ vapor is significantly stronger than that from $D_2O$ vapor.

$H_2O$ and $D_2O$ monomers have virtually identical ionization energies [15], so within the simplest free-electron plasma scenario, they should not produce THz emission with any significant isotope effect from photoionization alone. However, isotope effects do occur in the THz absorption spectra of water vapor due to monomer rotational-vibrational excitations and in discharge plasmas have been attributed to the dissociative recombination of hydronium-ion water clusters [16]. Water clusters, especially hydronium-ion protonated [13] and deuterated [16] clusters, $(H_2O)_nH^+$ and $(D_2O)_nD^+$, known to occur naturally in water vapor [9-13], might contribute to isotope-dependent THz emission because of their remarkable temperature stabilities [11] and unique THz vibrational modes (Fig 4c,d). Among these, the pentagonal dodecahedral $(H_2O)_{21}H^+$ cluster is one of the most stable "magic-number" cluster ions observed spectroscopically in water vapor [17,18]. We use this example within a molecular-orbital and vibrational normal-mode scenario to propose a simple model for isotope-dependent THz emission by optically pumped water cluster ions.

Fig. 4 shows the density-functional ground-state molecular-orbital energies and vibrational modes of the $(H_2O)_{21}H^+$ cluster, which from molecular-dynamics simulations are qualitatively unchanged well above 100 degC, where the clusters remain remarkably intact. Calculations for other $(H_2O)_nH^+$ clusters known to be present in ambient water vapor [9-13,17,18] possess similar manifolds of THz vibrations, and like Fig. 4a for $(H_2O)_{21}H^+$, their HOMO-LUMO energy gaps correspond roughly to the applied 400 nm laser wavelength. However, the numbers of molecular orbitals and vibrational modes per cluster increase dramatically with increasing n. Density-functional calculations for *neutral* water clusters such as pentagonal dodecahedral $(H_2O)_{20}$ have also been performed, yielding results very similar to Fig. 4. Molecular-dynamics studies of the $(H_2O)_{20}$ cluster have produced bands of cluster vibrations in the 0-1000 $cm^{-1}$ range similar to those shown in Fig. 4c [19]. Common to *all* the water clusters studied are: (1) lowest unoccupied (LUMO) energy levels like those shown in Fig. 4a, which correspond to the huge, delocalized "*S*"-, "*P*"-, "*D*"- and "*F*"-like cluster "surface" orbital wavefunctions shown in Fig. 4b; (2) a narrow band, like that shown in Fig. 4c, of "surface" vibrational modes between 1.5 and 6 THz (50 to 200 $cm^{-1}$) due to O-O-O "bending" motions between adjacent hydrogen bonds. Surface O-O-O bending vibrations of large water clusters in this energy range have been observed experimentally [20]. The vectors in Fig. 4d represent the directions and relative amplitudes for the 1.5 THz mode of the hydronium ($H_3O^+$) oxygen ion coupled to the O-O-O "bending" motion of the $(H_2O)_{21}H^+$ cluster "surface" oxygen ions. *Near-ultraviolet* excitation of an electron from the HOMO to LUMO (Fig 4a) can put the electron into the bound "*S*"-like cluster molecular orbital mapped in Fig 4b. Occupation of the LUMO produces a stable state of such a water cluster, even when an extra electron is added, the so-called "hydrated electron". In contrast, a water monomer or dimer has virtually no electron affinity. *Near-infrared* absorption can further excite the cluster LUMO "*S*"- like electron into a higher LUMO such as the "*P*"-like cluster orbital in Fig. 4b. *Infrared* excitations within the LUMO manifold can then decay vibrationally according to the Franck-Condon principle. The lowest-frequency water cluster vibrations induced by optical pumping should be the 1.5-6 THz "surface" modes of Fig. 4c,d. They span the same frequency range as the measured water vapor THz wave amplitude shown in Fig. 3b.

The calculated ground-state dipole moment of the $(H_2O)_{21}H^+$ cluster is nearly 10 Debyes, as compared with the 1.86 D moment of water monomer. The THz vibrations induced by optical pumping produce large oscillating electric dipole moments, even exceeding 10 D. Since the power radiated from an oscillating dipole is proportional to the square of the dipole moment, an optically excited water cluster ion like $(H_2O)_nH^+$ is a potentially strong source of THz radiation. Ambient water vapor containing $10^{17}$ cm$^{-3}$ water monomers typically contains only $10^4$ to $10^6$ $(H_2O)_nH^+$ clusters cm$^{-3}$, whereas neutral water dimer and cluster concentrations are as high as $10^{10}$ to $10^{12}$ cm$^{-3}$ [9,13]. Nevertheless, such isolated $(H_2O)_nH^+$ clusters can in theory be pumped *collectively* by photons into *resonant* vibrational (phonon) modes of lowest frequencies 1.5-6 THz, analogous to a Bose-Einstein condensation, acquiring giant electric dipoles of the order of $10^5$ to $10^7$ D cm$^{-3}$ [22]. This greatly enhances the possible contribution of water clusters to radiation absorption and emission and has been argued to explain the strong far-infrared and submillimeter absorption of radiation in Earth's atmosphere [13,19,22]. The emission amplitude of an oscillating electric dipole is inversely proportional to the vibrating atomic masses, suggesting that THz emission from water vapor due to optically pumped $(H_2O)_nH^+$ clusters should be larger in amplitude than that from $(D_2O)_nD^+$ clusters. The *differences* between the light and heavy water vapor THz emission amplitudes of Fig. 3a are qualitatively consistent with this scenario. However, isotopic differences in rovibrational absorption by the predominant monomers, including those produced in excited vibrational states by the dissociative plasma recombination of hydronium-ion water clusters [16], may be primarily responsible or at least relevant. Because of the expected relatively low concentration and variety of water cluster ions that occur naturally in the injected water vapor and in the cell, it was not possible to detect in the THz emission the vibrational signatures of specific water clusters such as $(H_2O)_{21}H^+$. Future THz emission experiments on water vapor should include the combined use of cluster size selection and mass spectrometry [9,11,18,19]. Since our model predicts optically induced water cluster surface O-O-O bending modes [20] at 1.5-6 THz (Fig. 4c,d), future THz emission experiments should search for a significant oxygen isotope effect using $H_2^{18}O$ vapor.

**References**

1. R.T. Hall, D.Vrabec, J.M. Dowling, Appl. Opt. 5 (1966) 1147.
2. P. Gaal, et al., Phys. Rev. Lett. 96 (2006) 187402.
3. D.J. Cook, R.M. Hochstrasser, Opt. Lett. 25 (2000) 1210.
4. M. Kreß, Opt. Lett. 29 (2004) 1120.
5. T. Bartel, Opt. Lett. 30 (2005) 2805.
6. M. Kreß, et al., Nature Physics 2 (2006) 327.
7. X. Xie, J. Dai, X.-C. Zhang, Phys. Rev. Lett. 96 (2006) 075005.
8. K.Y. Kim, et al., Optics Express 15 (2007) 4577.
9. H.R.Carlon, C.S. Harden, Applied Optics 19 (1980) 1776.
10. F.N. Keutsch, R.J. Saykally, Proc. Natl. Acad. Sci. USA 98 (2001) 10533.
11. M. Tsuchiya, T.Tashiro, A. Shigihara, J. Mass Spectrom. Soc. Jpn. 52 (2004) 1.
12. H. Harker, et al., J. Phys. Chem. 109 (2005) 6483.
13. K.L. Aplin, R.A. McPheat, J. Atmos. Solar Terrest. Phys. 67 (2005) 775 .
14. Y. Chen, et al., Appl. Phys. Lett. 91 (2007) 251116.
15. R.H. Page, et al., J. Chem. Phys. 88 (1988) 2249.
16. E.A. Michael, et al., J. Molec. Spect. 208 (2001) 219.
17. J.-W. Shin, et al., Science 304 (2004) 1137.
18. M. Miyazaki, et al., Science 304 (2004) 1134.
19. I. P. Buffey, W. Byers Brown, H.A. Gebbie, J. Chem. Soc. Faraday Trans. 86 (1990) 2357.
20. J. Brudermann, P. Lohbrandt, U. Buck, Phys. Rev. Lett. 80 (1998) 2821.
21. J.R.R. Verlet, et al., Science 307 (2005) 93.
22. H.A. Gebbie, Nature 296 (1982) 422; Science Progress 78 (1995) 147.

**Figure Captions**

**Fig. 1.** Schematic diagram of the experimental arrangement. **(a)** a vapor glass cell or **(b)** a pulsed jet nozzle.

**Fig. 2.** Comparison of the THz wave generation amplitudes of water vapor, several noble gases, and nitrogen gas at 20 Torr.

**Fig. 3.** THz wave generation amplitude (in arbitrary units) from water vapor, heavy water vapor, and nitrogen. **a.** Pressure dependence. For the pressure less than 20 Torr, the data were taken from the cell, while at higher pressure (>20 Torr), the data were taken from the pulsed nozzle. **b.** Frequency dependence for a range of pressures.

**Fig. 4.** Ground-state density-functional molecular orbital states and vibrational modes of the $(H_2O)_{21}H^+$ cluster. **a.** Cluster molecular-orbital energy levels. The HOMO-LUMO energy gap is 3eV. **b.** Wavefunctions of the lowest unoccupied cluster molecular orbitals. **c.** Vibrational spectrum. **d.** 1.5 THz vibrational mode. The vectors show the directions and relative amplitudes for the oscillation of the hydronium ($H_3O^+$) oxygen ion coupled to the O-O-O "bending" motion of the cluster "surface" oxygen ions

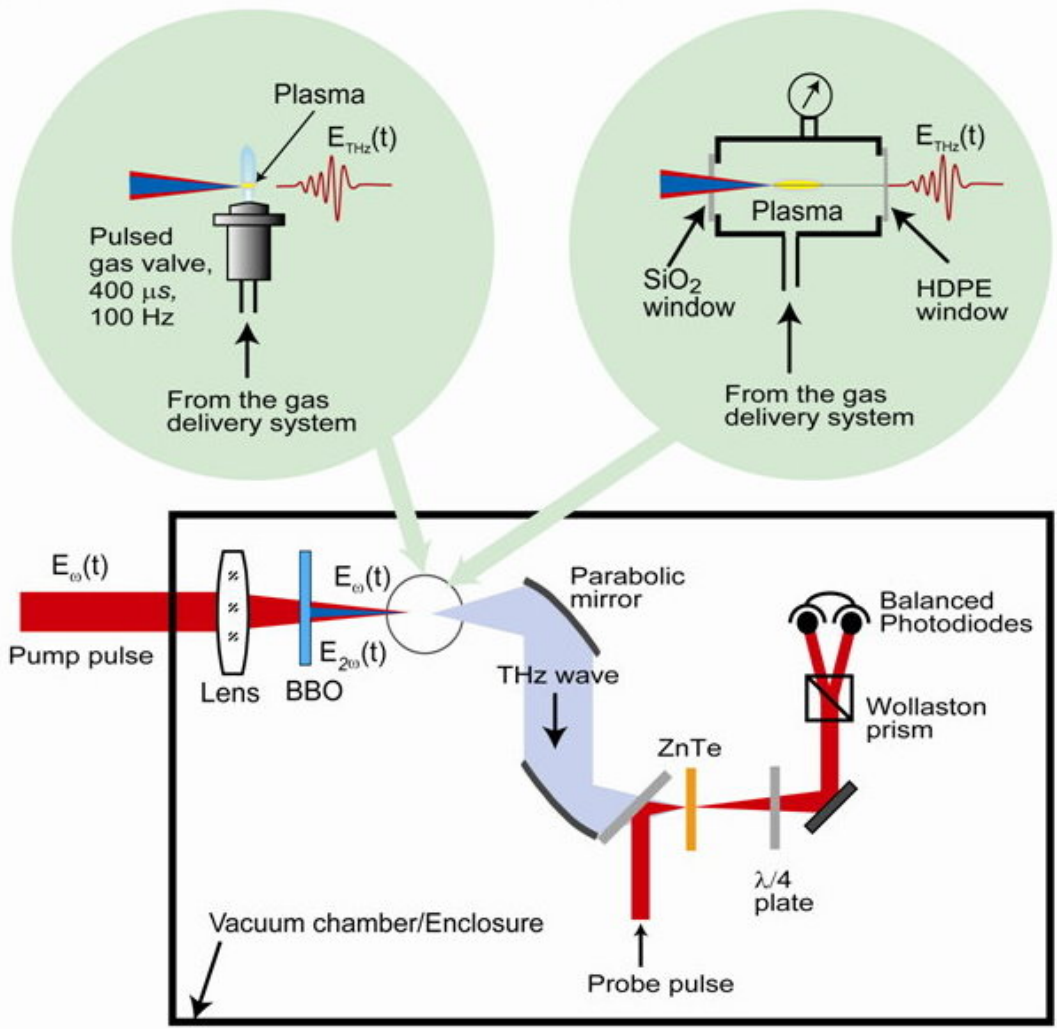

**Fig. 1**

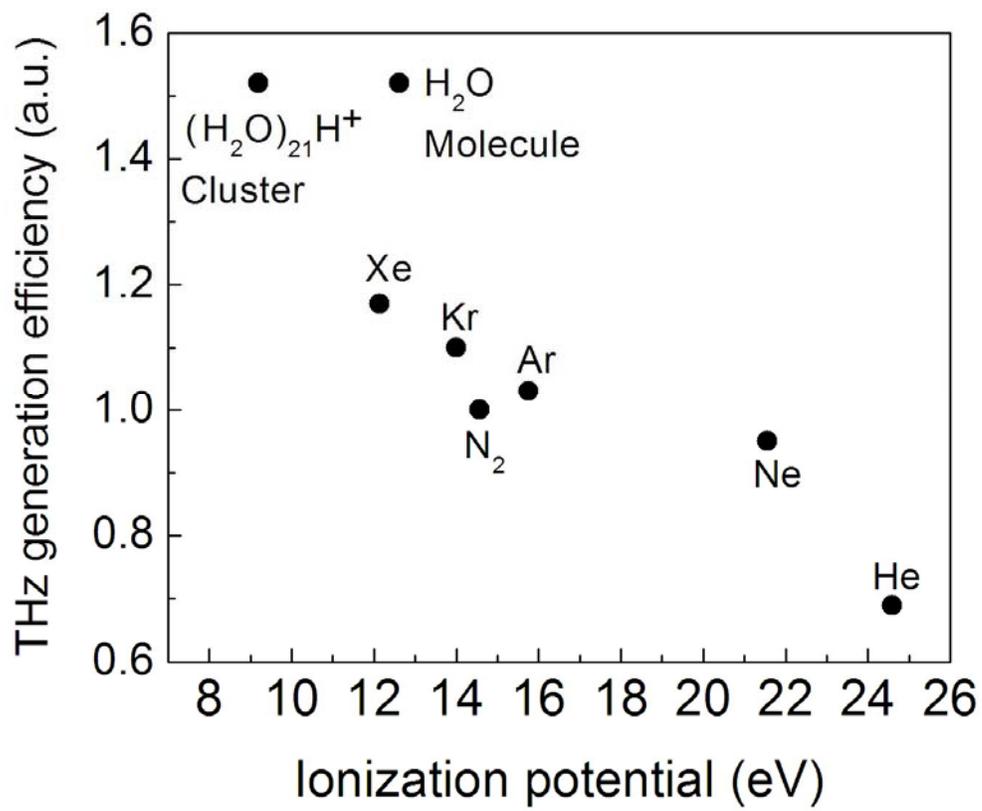

**Fig. 2**

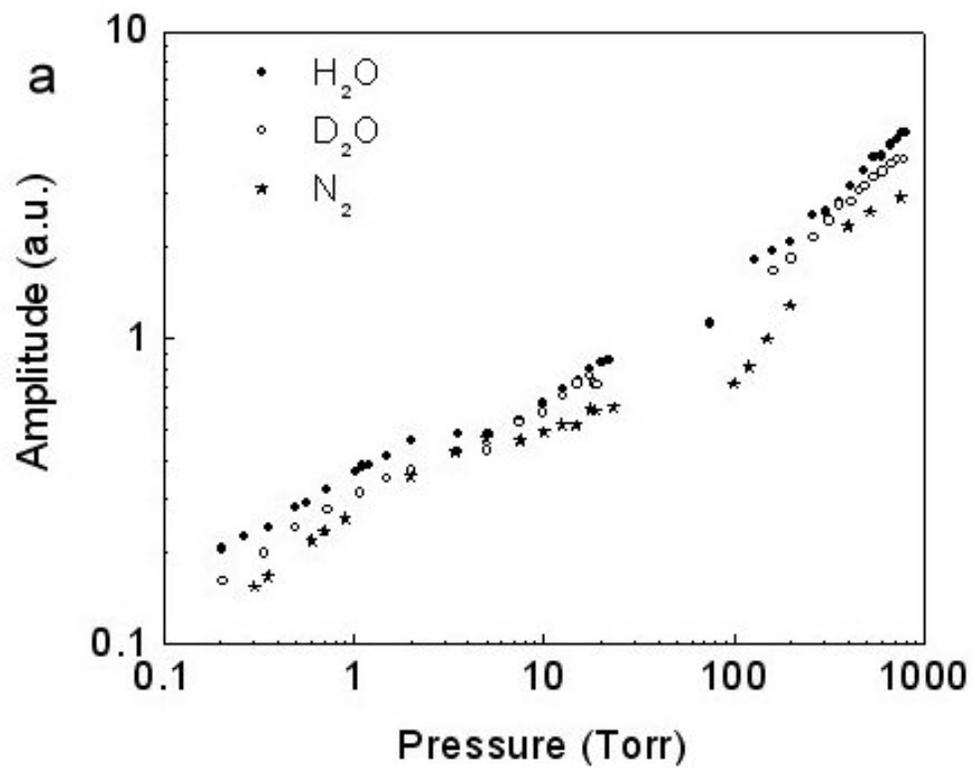
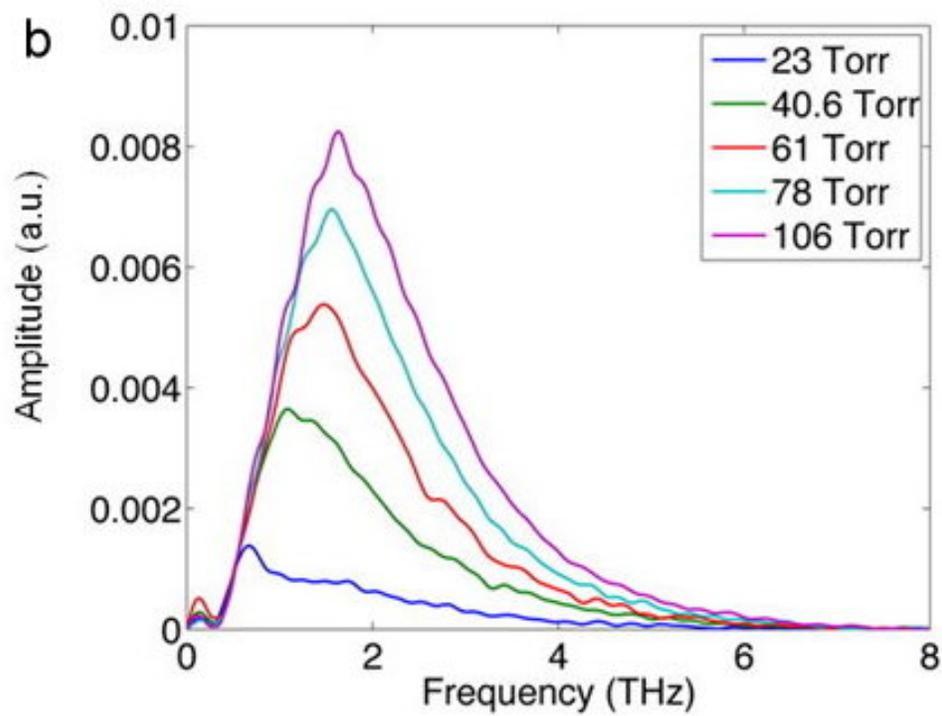

**Fig. 3**

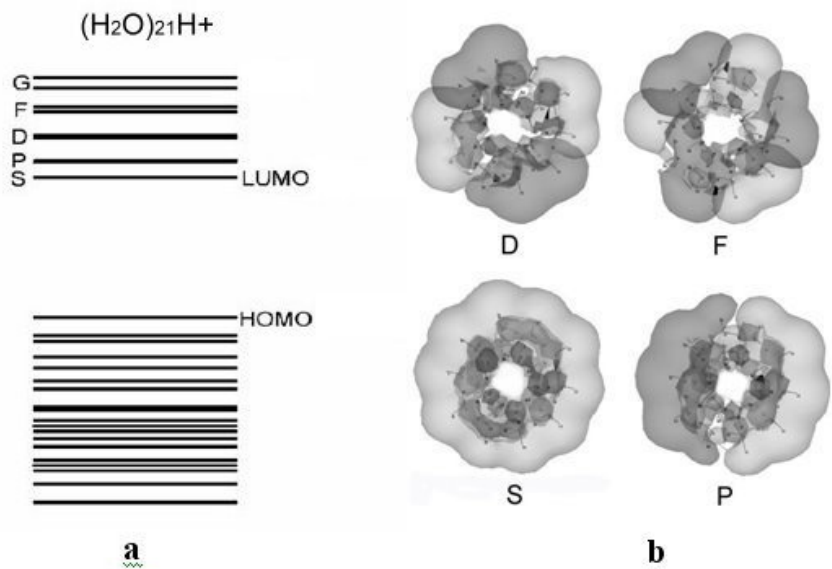
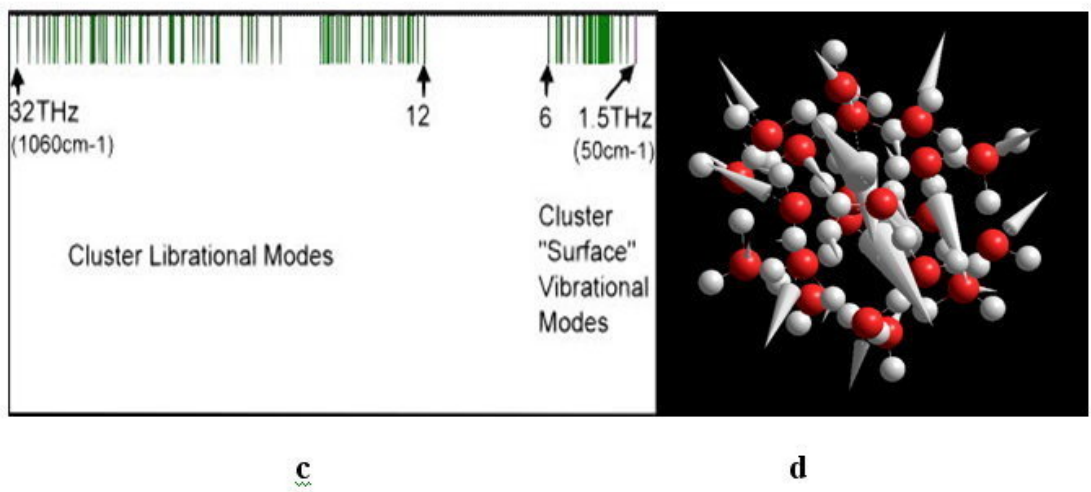

**Fig. 4**